# Architecting Time-Critical Big-Data Systems (preprint)

authorsegmentP. Basanta-Val, N. C. Audsley, A. J. Wellings, I. Gray, N. Fernández

*Abstract*— Current infrastructures for developing big-data applications are able to process –via big-data analytics- huge amounts of data, using clusters of machines that collaborate to perform parallel computations. However, current infrastructures were not designed to work with the requirements of time-critical applications; they are more focused on general-purpose applications rather than time-critical ones. Addressing this issue from the perspective of the real-time systems community, this paper considers time-critical big-data. It deals with the definition of a time-critical big-data system from the point of view of requirements, analyzing the specific characteristics of some popular big-data applications. This analysis is complemented by the challenges stemmed from the infrastructures that support the applications, proposing an architecture and offering initial performance patterns that connect application costs with infrastructure performance.

## I. INTRODUCTION

Current trends in computer science refer to *big-data systems* to describe large and complex data centric applications that cannot be run properly with current data processing tools [1-6][50-52]. These tools are insufficient to analyze, capture, cure, search, store, transfer, or visualize the amount of data processed by this type of applications [46-47]. Big-data systems might also refer to algorithms that perform some type of analytics [15, 28-29] that extract valuable information from data, to find new correlations, to spot business trends, or to combat crime.

Big-data scenarios [29-30] are characterized by the existence of an enormous amount of information (coming from mobile devices, logs, sensors, cameras, etc.) that needs to be processed to achieve a goal. A single PC cannot typically process this amount of data; thus, it is often processed by algorithms that run in hundreds or thousands of servers hosted in private clusters. In many cases, they can also use the Internet, which offers cheap hosting to big-data storage and computational applications via cloud computing infrastructures, as a cost-effective solution.

From a business perspective, big-data offers an opportunity to increase operational efficiency in an enterprise [5-7]. It may reduce operational costs by detecting inefficient policies which may be replaced by more efficient ones. Also, it may discover new business niches and opportunities, mining information available into the organization.

Time-critical systems [7-8] refer to systems subject to certain temporal restrictions, which typically consist of maximum deadlines for an input event to be processed, and/or an output to be generated. These maximum deadlines are derived from the characteristics of an external environment that imposes physical requirements on applications. In time-critical systems, response-times can be in the order of milliseconds, or microseconds, e.g. for system control; they can also be longer, e.g. when interacting with human interfaces. Typical time-critical systems [29-34] have benefited from different general-purpose computational infrastructures (e.g. programming languages, operating systems, modeling languages, etc.) including specific computation algorithms that take advantage of the application characterization to estimate, a priori, maximum response times. Currently, there is not a clear definition of how time-critical and big-data systems should be merged to produce a "time-critical big-data system". However, there are some pioneering research initiatives [9-11] that seem to identify different opportunities stemmed from the combination of these two types of systems.

Implicitly, many big-data applications have requirements in terms of maximum expected deadlines that have to be satisfied by the underlying infrastructure. In these cases, it seems that many of the existing techniques for time-critical systems may be beneficial for the time-critical big-data applications to estimate maximum deadlines for running their analytics. Work like [9-11] shows the relationship among the response-time of the applications and their performance, establishing different mathematical models that connect worst-case response-times and the number of machines with quasi-linear formulations.

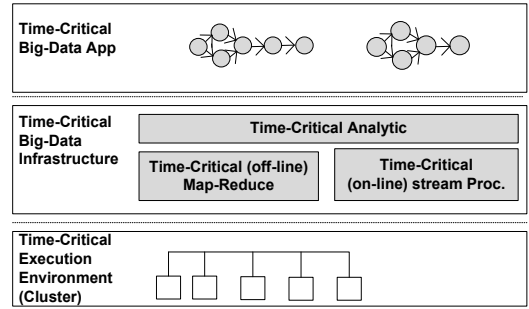

Fig. 1. Holistic Time-Critical Big-Data System

The contribution of this paper is an architecture for time-critical big-data systems (summarized in Fig. 1) able to run time-critical analytics. These analytics are supported by an infrastructure able to run different programming abstractions on a large number of machines, arranged as a cluster. To implement the architecture, our approach is to extend Spark [23] (to support time critical map-reduce), and an Apache Storm [16] with a backward time-critical stack. Also part of our contribution is a set of techniques that profile the performance of a time-critical infrastructure.

The rest of the paper is organized as follows. Section II deals with the application domains: from science applications to social networking, with real-time requirements. Those domains raise a number of issues or technical challenges (Section III). Section IV introduces an architecture for big-data applications standing on common off-the-shelf technologies (Apache Storm [16] and Spark [23]), which are equipped with time-critical engines. Section V introduces specific technological mappings and application use-cases. Section VI introduces evaluation results for both stacks in two types of applications. Section VII describes the related work and Section VII summarizes results and exposes our on-going work.

## II. DOMAINS

The exploration of the different initiatives for time-critical big-data systems starts by analyzing applications taken from different use-cases from science, data processing, financial services, and social networks. The goal of this section is to illustrate the time-critical characterization of these applications.

### A. Science Applications

There are two classical examples of time-critical big-data science applications: the Large Hadron Collider (LHC) [12], and the Square Kilometer Array (SKA)[22]. LHC experiments involve 15 million sensors delivering data 40 million times per second. From a total of 600 million collisions per second, only 100 collisions per second are interesting. Thus, the collider outputs a 300 Gb/s stream that has to be filtered to 300 Mb/s for storage and later processing.

Implicitly, in LHC all this data cannot be stored in its raw format and has to be processed online, with computational units running at 100% of their computational speed, before being stored. Working with less than 0.001% of the sensor data, the data flow from all experiments represents 25 Petabytes. In a case where all data is to be recorded, the resulting flow would exceed 150 million Petabytes of data.

### B. Data Aggregators

Another type of applications related to big-data systems that aggregate data from different sources, known as data aggregators [13]. Their aggregation phases refer to data coming from the Internet and also data coming from other private infrastructures, like sensors in cars, buses, and planes. This data is explored by the big-data infrastructure to look for different patterns globally. As in the previous case, in most scenarios the time-critical requirement refers to the possibility of using enough computation facilities to be able to process the data without overflowing. In addition to this stability condition, another set of deadlines may be established to perform specific detections.

### C. Financial Transaction

Online and offline business analytics [14] represent another potential application scenario of time-critical big-data systems. Online analytics (like those used in High Frequency Trading) define the strategies used by auctioneers to decide if a stock is to be sold or bought. In this type of application, the shorter the response time of the intelligent system, the better the results are. In the case of offline analytics, they may also process data stored in logs/data servers to predict the future behavior of certain markets or stocks by using well-known techniques. As in the online strategy, a delayed response may result in financial losses.

### D. Social Networking

Social networks are another source of big-data information. They produce a myriad of streaming dynamic data that may be processed, for instance, to offer intelligent advertising to end-users [14]. These types of applications are difficult to model as time-critical systems, because they tend to have very complex data sources and worst-case computation times are difficult to calculate. In many cases, it seems that the infrastructure may benefit from having some type of adaptive quality of service algorithm to manage the nodes of a cluster, according to the application requirements. In specific cases it seems possible to establish other types of operational deadlines on the analytics. This is the case, for instance, of event detection (like earth-quake detection [45]) via social networking, where a sub-second scale deadline may be established as a requirement.

## III. TECHNICAL CHALLENGES

Another way to approach time-critical big-data systems is to analyze the infrastructures currently given to practitioners, studying the technical challenges which stem from trying to implement time-critical applications using these infrastructures. In many cases these infrastructures have not been properly adapted to the requirements of time-critical applications, giving rise to new requirements for developing these systems.

### A. Lack of Time-Critical Big-Data Facilities

Most big-data infrastructures were designed with general-purpose applications in mind and are silent on the requirements of time-critical systems. All main development platforms based on Hadoop [15] and map-reduce target HPC platforms and do not support the idea of defining deadlines. Only in some specific cases [1], like the case of the Apache Storm distributed stream processor [16], do they seem to target online computation. But even in these cases, most infrastructures seem to be focused in general-purpose applications rather than considering time-critical performance.

Among the list of issues that have to be considered when designing an effective time-critical infrastructure for big-data applications, one should address the following aspects:

- *Ability to process high volume of data*: Among the first features required from a big-data system is the ability to process a huge amount of data across multiple computational nodes. This capacity refers not only to the number of computational nodes in the system, but also the amount of disk storage, memory, and communication resources available among the different nodes of the cluster. Current algorithms are focused on general purpose computation [9].
- *Distribution and Parallel Computation*: Most big-data applications are distributed applications where several connected nodes share information and resources to accomplish a mission, typically, to perform big-data analytics. Thus, many infrastructures are able to distribute the computation among different nodes of a cluster, using different policies. However, in most cases, the algorithms used do not take into account the time-critical nature of the application, producing setups that are non-optimal from the perspective of the application.
- *Data Locality*: Another important source of non-determinism that may impact the response time of the applications is the availability of local or remote access to data. Remote access to data tends to introduce a response time penalty of more than an order of magnitude in some big-data applications, in addition to an increased jitter (see [20]). However, remote communication opens the door to introduce parallel and distributed computing to reduce computation times, which may be used to effectively reduce application worst-case response times.
- *Fault model*: Big-data applications run in faulty infrastructures where computation nodes may appear and disappear dynamically. This introduces additional requirements for the infrastructure, which has to provide predictable recovery policies ([20-21]).

In addition to the shortcomings in the infrastructure, there is also a general lack of models for time-critical big-data. Current existing tools for big-data (e.g. those using model driven architecture approaches) have been designed for general purposes and are silent on the identification of specific optimizations for time-critical applications. Thus, these tools need to be extended with time-critical characteristics, required for developing time-critical big-data applications.

### B. Time-Critical Algorithms and Analytics

In most cases, big-data applications are composed of a set of algorithms that run on data, trying to perform some type of analysis; such as a prediction of trends and patterns [17]. Typically, the analytics carried out may be classified into three groups: descriptive, predictive and prescriptive. Descriptive analytics have the goal to condense big-data into smaller chunks. Predictive analytics take data and produce as output a prediction regarding future values or behavior of that data. Another interesting characteristic of some analytics is that they may improve their output quality as the amount of computation time or data increases. This property may be used to produce time-critical efficient analytics that tradeoff output quality and computational cost. In many cases, the computational model of a particular class of analysis may be expressed as a directed acyclic graph (DAG) [9]. This DAG may be used to compute a worst-case response-time for the analytic, provided that the worst-case execution time for all stages in the DAG is known.

### C. Security and Privacy

Another major concern of a big-data system is the existence of a secure and private execution environment [18]. Among the challenges imposed to a big-data system are: the definition of secure computations in distributed programming frameworks, secure data storage, transaction logs, endpoint validation, real-time security, data centric security, granularity of access-controls, audits and data provenance.

The proposed time-critical data system addresses the basic challenge of producing a time-critical infrastructure for the computational models of Spark and Storm. In the improved architecture, analytics are driven with priorities. Section IV contains the portable architecture and the computational framework. Section V deals with software aspects regarding an implementation on top of Spark and Storm. Lastly, Section VI describes performance aspects of the computational framework. All other challenges described in the section have been set aside.

### IV. TIME-CRITICAL BIG DATA SYSTEM

In this section, we explore an architecture defined to support big-data systems. The proposed architecture is based on existing technologies, which are enhanced with a time-critical dimension. In essence, the model mixes two main technologies: Apache Storm [19] which is an efficient approach for sub-second delays, and Apache Spark [24], which offers a rich and an efficient framework on which to implement offline map-reduce applications. The model is also highly inspired on previous architectures described in the context of distributed real-time systems for Java [20][21][25][53]. From this distributed context, the architecture also takes specific support for parallel processing [9], and several online scheduling mechanisms for predictable reconfiguration strategies (see [26]).

### A. Overview

The architecture proposes a layered approach for big-data, combining traditional time-critical middleware [30] with big-data architectures [1][31]. Our reference model consists of four main layers, each one of them dealing with different aspects of the big-data system (see Fig. 2):

- **Applications**: At the top of the architecture are applications, which are named analytics. They are arranged as a directed acyclic graph structure, easing deployment in a large cluster of machines. The main type of consideration that has to be taken into account is their nature (see Fig. 3):
- *Time-critical* (deadline oriented): The main requirement imposed on this layer is the existence of time-critical requirements. Typical requirements are deadlines, which range from the sub-second range to seconds, days, weeks, etc. These deadlines on the analytic are essential to determine the number of machines required to perform the analytic. In a harmonious and balanced system, the number of machines required to meet a deadline increases as the amount of data does (linearly).
- Integrate *offline and online analytics*: Typically, analytics can be classified as offline and online. Offline analytics (sometimes named batch analytics) tend to explore a large volume of data and they have large CPU consumption models. On the other hand, online analytics have shorter deadlines which require the use of different techniques to meet time-critical deadlines.

- **Tools**: Supporting applications are tools, which define the second level in the architecture. These tools satisfy different aspects of an analytic. In the particular case of a time critical system, they offer support to deadlines and any other type of requirements. To satisfy deadlines, tools have fine control on the resources available to machines, which have to be properly controlled and configured. The types of resources managed include disk bandwidth, memory, CPU Cores, and communications.

Different types of applications require different types of resources to be controlled. In the case of online applications,

they are likely to deal with lower data volumes, and higher speeds. On the other hand, offline infrastructures deal with huge data sets and, a priori, do not have a direct interaction with end-users.

- **Infrastructure**: Infrastructure refers to fundamental facilities servicing applications and implementing necessary blocks for running analytics. At this layer, there are at least five different types of sub-infrastructures, whose proper management may have a marked impact on performance and on the response time of applications, including:
  - *Cluster management support*: Many of the different tools use common facilities to access a cluster. They use different facilities from local or remote nodes, which are accessed via specific operating system interfaces or/and low level managers for storing computational models and/or connection managers.
  - *Operating system (OS)*: Typically, an OS is in charge of controlling a set of resources via a well-defined interface. In some cases, interfaces include resource managers that control resources assigned to an application. Some specific OSs define policies for real-time performance, that can be enhanced to add time-criticality.
  - *Storage managers*: In close connection with operating systems, there are storage managers, which control different storage systems. In some cases, they offer support for different quality-of-service policies, which can be exploited in time-critical systems.
  - *Node managers*: They control a set of machines remotely from a manager. Typically, those node managers handle different classes of machines, with similar installed software and hardware stacks.
  - *Connection managers*: They control communications, and are crucial as the communication needs and the cluster increase in size.

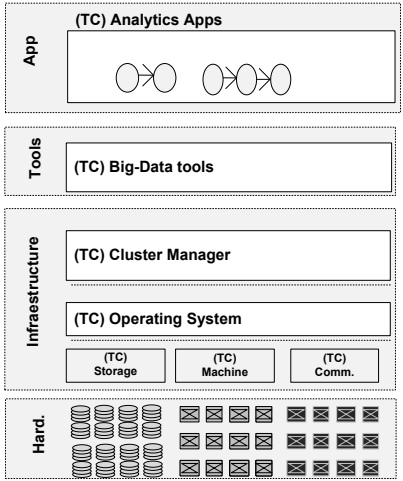

Fig. 2. Architecture for Time-Critical Big-Data

- **Hardware**: At the bottom of the architecture is hardware. In many big-data systems, typical hardware layers are arranged as clusters of quasi-identical machines (potentially virtual machines). In a more general framework, the main facilities to be considered at this layer are:

1. Computation: This refers to different computation resources (CPUs, cores).
2. Storage: This refers to storage units required to process huge volumes of data.
3. Communication: This refers to different facilities that interconnect computational resources and/or storage units.

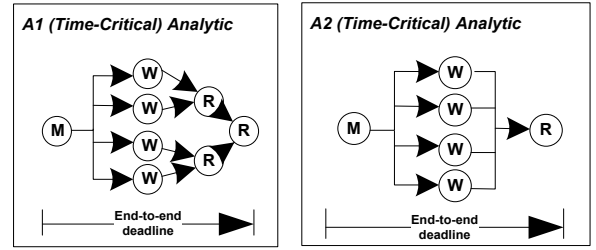

Fig. 3. Time-critical analytics: Each analytic describes a simple end-to-end time-critical constraint. In the depicted scenarios, the work (W), executed in parallel by an analytic, is preceded by a map (M) phase and followed by a reduce (R) phase. The end-to-end requirement is described as an analytic deadline.

### B. System Model

The model described in Fig 3, has been formalized using real-time systems theory to produce a time-critical characterization, useful for map-reduce and distributed stream processing. The essence of the technique consists in splitting different stages in applications, which are scheduled locally with other analytics and that interact with other parallel (||) and sequential stages ($\rightarrow$).

#### 1) Time-Critical Analytics

From a formal perspective, a time-critical big-data system ($TC\_BDS$) is composed of a set of $n$ analytics ($TC\_A_i$):

$$TC\_BDS \stackrel{\text{def}}{=} (TC\_A_1, \dots, TC\_A_n) \quad \text{(eq.1)}$$

Where each analytic ($TC\_A_i$) is represented by an execution graph ($TC\_DAG_i$) and a set of performance requirements ($TC\_RQ_i$):

$$TC\_A_i \stackrel{\text{def}}{=} (TC\_RQ_i; TC\_DAG_i) \quad \text{(eq.2)}$$

The execution graph models are commonly used in many systems. For instance, they are the computational model in Apache Storm and they can be also used to model map-reduce interactions of Apache Spark.

Our basic requirement is a deadline for the whole analytic ($D$). Deadlines are defined by the analytic and may include data reading, writing, blockings, and any other logic associated with the execution of the analytics:

$$TC\_RQ_i \stackrel{\text{def}}{=} D_i \quad \text{(eq. 3)}$$

Likewise, each direct acyclic graph ($DAG_i$) is composed of a set of $stages(i)$. In each stage, the characterization is defined with a minimum inter-arrival time ($T_i^j$), a partial deadline for each stage ($D_i^j$), and a worst-case execution ($C_i^j$) associated to the stage:

$$TC\_DAG_i = \begin{pmatrix} [T_i^1, \dots, T_i^{stages(i)}] \\ [D_i^1, \dots, D_i^{stages(i)}] \\ [C_i^1, \dots, C_i^{stages(i)}] \end{pmatrix} \quad (eq.\ 4)$$

*2) Cluster Infrastructure*

In the architecture, the infrastructure runs on a cluster (TC_CLS) of $m$ machines ($\pi$):

$$TC\_CLS \stackrel{def}{=} (\pi_1, \dots, \pi_m) \quad (eq.\ 5)$$

Each machine offers a normalized maximum utilization ($U_k$) to the system and it also introduces a maximum blocking time ($B_k$), which typically affects the application and refers to the maximum time an application may be awaiting a resource:

$$\pi_k \stackrel{def}{=} (U_k, B_k) \quad (eq.\ 6)$$

Blockings are very useful to model non-preemptive behaviors, related to packet scheduling and also the effects of priorities – e.g. high priority process blocked by a lower priority process holding a required resource. They are also applicable when modeling access to disk and network facilities.

*3) Time-Critical Characterization*

To be properly characterized, each stage of an application should be assigned to a machine of the cluster ($\pi_i^j$). To help the application, the system has to declare a priority ($P_i^j$) which has to be honored in all nodes. In addition, in choosing a node, the analytic of a stage also suffers a blocking ($B_i^j$) from the infrastructure.

$$TC\_DAC_i = \begin{pmatrix} [T_i^1, \dots, T_i^{stages(i)}] \\ [D_i^1, \dots, D_i^{stages(i)}] \\ [C_i^1, \dots, C_i^{stages(i)}] \\ [\pi_i^1, \dots, \pi_i^{stages(i)}] \\ [P_i^1, \dots, P_i^{stages(i)}] \\ [B_i^1, \dots, B_i^{stages(i)}] \end{pmatrix} \quad (eq.\ 7)$$

After describing all different analytics from the application as fully configured sets of stages, one may determine bounds for the response-time of an analytic. This type of configuration is intended to be used with worst-case computational models for distributed and parallel computing.

*4) Meeting analytic deadlines: Schedulability analysis*

Once the system is fully configured, using scheduling theory one may check if all deadlines in applications are met. Typically, one may resort to a general formulism, which is exact and valid for all different tasks. And also, there is a utilization bound (valid with T=D) constraint and priorities assigned inversely to periods; i.e. shorter periods have higher priorities). These two techniques are the core of the contribution of the article.

The first equation refers to the response time (TC$_{RTi}$) associated to each different segment of an application. Taking as a starting point the characterization given in Eq. 1- 7, one may calculate the worst-case response time ($TC\_RT_i^j$) in each node, using the following recursive formulation:

$$TC_{RT_i^j} < B_i^j + C_i^j + \sum_{\forall (z)\ in\ HP(i,j)\ and\ \pi(i,j)} \left\lceil \frac{TC_{RT_i^j}}{C_z} \right\rceil C_z \quad (eq.\ 8)$$

Basically, the formalism applied is based on the response time analysis (RTA) [31]. RTA considers that the worst-case time of a segment is equal to the required execution ($C_i^j$), plus the blocking ($B_i^j$) experienced from the infrastructure plus the interference of those tasks with higher priorities ($HP(i,j)$) hosted in the same node $\pi(i,j)$. Each of these tasks introduce a ($C_z$) extra demand every $T_z$ time units. To solve all (eq. 8) equations in all nodes, one needs an incremental method to calculate the right response time, which has polynomial complexity.

The second type of constraints refers to the extra conditions that enable calculations of the worst-case response times. Segments are grouped with two compositions: sequential ($\rightarrow$) and parallel ($\|$). For each of them, one needs to define worst-case response times for the connection of different segments. Eq. 9 refers to the extra conditions one has to calculate for a sequential interconnection and Eq. 10 for a parallel one.

To calculate the worst-case computational times of two sequential ($\rightarrow$) stages ($i$) and ($i'$), the worst case scenario is to add partial contributions of two individual segments:

$$TC_{RT_{i \rightarrow i'}^{j \rightarrow j'}} < TC_{RT_i^j} + TC_{RT_{i'}^{j'}} \quad (eq.\ 9)$$

In addition to that, there is parallel relationship ($\|$). If two stages run in parallel, then the response time of its aggregation is the maximum of the worst-case response time of any of them:

$$TC_{RT_{i\|i'}^{j\|j'}} < \max(TC_{RT_i^j}, TC_{RT_{i'}^{j'}}) \quad (eq.\ 10)$$

Another type of techniques, which can be only used in specific cases (with T=D) and priorities assigned inversely to their periods) is to offer a global utilization bound that can be computed in linear time (i.e. linearly with the number of elements to be computed). In this work, we refer to a specific formalism of multiprocessors that establishes a safe bound for the maximum number of resources [33, 34]. It connects the number of machines required to support the system (when the deadline is equal to the period: $T=D$ for rate monotonic assignments). The framework extends this model with blockings. The formalism connects the total utilization of the application ($U_{TC\_BDS}$) with the number of required nodes ($m$) in the cluster to meet different application deadlines.

Assuming that all deadlines in all segments are increased by the blocking time ($B_i^j$):

$$T_i^j + B_i^j = D_i^j \text{ for all i, j}$$

And a maximum utilization provided by each node ($U_{max}$):

$$U_z >= U_{max} \text{ for all z in 1 .. m.}$$

then:

$$U_{TC\_BDS} = \sum_{\forall i,j} \left( \frac{C_i^j}{T_i^j} \right) < (m - 1/2) \cdot (U_{max}) \quad (eq.\ 11)$$

Notice, that blockings ($B_i$) reduce the response of an application, which gets increased (see Eq. 8) with the blocking. However, they do not have an impact on the number of resources of a utilization based system, because it only introduces delay. Another interesting result of the technique is that one may derive a safe bound for the maximum number of computational units ($m$) required for implementing the system. As the formulism shows, the number depends on the minimum period of the applications and the maximum activation costs.

This section provided the basic formulism used to check if the system is feasible or not. Enforcing a classical characterization (i.e. T, C, D, P, B) the approach introduces the equations used to derive worst-case computational models that may be iteratively computed and to obtain worst-case computational models. In addition to this iterative models, there are utilization bounds (with B+T=D) that offer sufficient bounds for the maximum utilization of the system.

## V. TIME-CRITICAL IMPLEMENTATION

The previous section defined a time-critical model for big-data. This section defines a complementary time-critical software stack, which is evaluated later in Section VI. The proposed model is based on a map-reduce cluster. This map-reduce cluster offers file-system management, which may host a large volume of data. It also includes support to manage a set of machines via cluster managers. The stack uses a common-off-the-shelf OS (Linux), which is installed in all nodes.

On top of this infrastructure there are tools that enable the possibility of performing analytics. Two used tools are Apache Storm [9], targeted at sub-second response times, and Spark [23] with enhanced map-reduce performance. Storm uses a distributed stream programming model, intended for online processing, while Spark is more commonly used in batch processing.

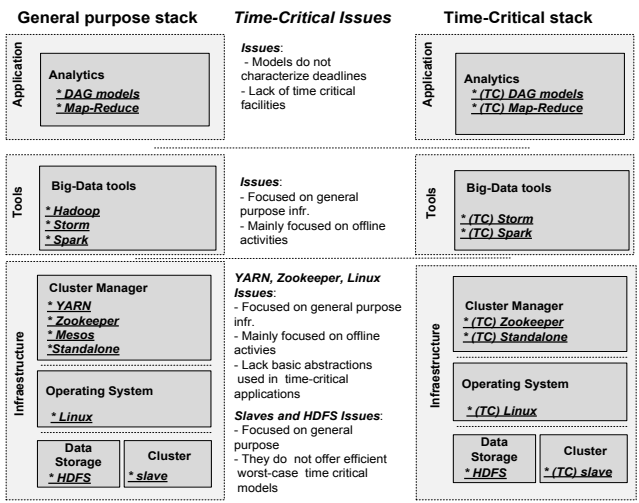

Fig. 4. Transforming a general purpose stack into a time-critical stack

Lastly, the uppermost layer offers the possibility of using map-reduce via Apache Spark or processing high-speed streams with Apache Storm. To be able to offer time-critical performance, the following mechanisms are required:
- A mechanism which assigns applications, according to their characterization and the previous described model. Typically, different resources managers (e.g. YARN, pluggable scheduler of Apache Storm, and the standalone cluster manager) have different policies that can be extended to take into account the nature of the analytic.
- A mechanism which performs resource reservations at the OS level. One common facility included in the Linux OS is `cgroups` [43], which enables aggregation of a set of tasks into hierarchical groups. Also, if they are properly configured, they help to enforce periodic activations in nodes.
- A mechanism to configure the priority of an application running on a cluster. This facility may be satisfied at application level or at cluster level.

### A. Distributed Stream Application Example

Our first example to show how a distributed stream processor works is based on a distributed word counter. The distributed word counter may be defined as a set of four stages, the first in charge of generating the flux of data, which is after that processed by a parallel counter that splits words contained in each message. After that, words are counted using a hash-table in another stage. Lastly there is an aggregation phase. Our distributed stream example consists of three stages which are arranged as a DAG (see Fig. 5).

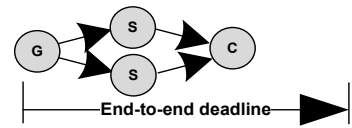

Fig. 5. Scenario under evaluation: Time-Critical Stream Counting. The application consists of one Spout generator (G) which feeds several splitters (S). All splitters feed a unique counter (C). There is an end-to-end deadline for the analytic, starting in the generator and ending up in the counter.

Now, let's consider the perspective of a programmer, providing an example from this perspective. This type of analytic consists of code that reads data with a generator stage (coded in `TC_Generator`, see Listing 1), a splitter (`TC_Splitter`, in Listing 2) which extracts words out of sentence, and a counter (`TC_Counter`, in Listing 3) that counts all words. The application to be deployed requires from a last class which creates the topology and allocates all the elements of the stream (`TC_Counter_App`, in Listing 4).

The generator uses a spout of Storm (see Listing 1), which has been modified to add time-critical information. The spout is a active class where the code invokes the `nextTuple()` method. The method is in charge of generating data that is sent to the next stage of the stream. This tuple is transferred in Line 17. In addition, the example includes a time-critical mechanism to configure enforcement properties. For instance, it is used to enforce the periodic activation of the stream in generation. It is also in charge of setting up a priority, which is enforced in all nodes.

Those tuples emitted by the spout go through the infrastructure and are processed by a bolt. In all bolts, it raises an invocation to the `execute()` method with the information of the tuple. In our particular case (see Listing 2), first the bolt checks if the stream is properly received (Line 08) with a call to the `tc_config()` method, which is also in charge of setting the priority and/or enforcing minimum applications interarrivals between different invocations. After the invocation, the stream is split (Line 09) into different pieces, which are sent to the last bolt. The last bolt (see Listing 3) is in charge of counting the words packed into a stream.

Lastly, to create the application, a topology is required which is then sent to the cluster (see Listing 4). The most important part is the one that connects different elements to build the topology, which defines an interrelationship among them. The generator is created and aggregated with `setSpout`. Then a bolt is created with a parallelism hint of 2 (Lines 10-12). The last element to be added is the bolt in charge of counting data (Lines 14-15). Lastly, the application is sent to the cluster of machines to be properly executed and the configuration file which contains the priorities and machines corresponding to all nodes (Line 18).

```
01:   public class TC_Generator implements IRichSpout
02:   {
03:     FileSource fs;
04:     OutputCollector collector;
06:     public void open(conf, context)
07:     {
08:     loadSource()
09:     storeConf()
10:     }
11:     public void nextTuple()
12:     {
13:       while(availableData())
14:       {
15:         tc_config();
16:         get_data();
17:         emitTuple();
18:       }
19:     }
20:     public void declareOutputFields()
21:     {  }
23:   }
```
Listing 1: Time-Critical spout. The `next_tuple` method call to `tc_config` to enforce an application defined behavior.

```
01:   public class TC_Splitter implements IRichBolt
02:   {
03:     public void prepare(conf, ctx)
04:     {  }
06:     public void execute(Tuple input)
07:     {
08:       tc_config();
09:       split();
10:       emit_data();
11:     }
12:   }
```
Listing 2: Time-Critical Splitter bolt. Prior to the execution of the application logic, it invokes a `tc_config()` method that configures the bolt according to the time-critical characterization.

```
01:   public class TC_Counter implements IRichBolt
02:   {
02:     Map counters;
03:     OutputCollector collector;
04:     public void prepare( conf, context)
05:     {  }
07:     public void execute(Tuple input)
08:     {
09:       tc_config();
10:       processTuple()
11:     }
12:     public void declareOutputFields(declarer)
13:     {
14:     }
15:   }
```
Listing 3: Time-Critical Counter Bolt. It does not emit any tuple. It only processes incoming tuples.

```
01:   public class TC_Stream_App
02:   {
03:     public static void main(String[] args)
04:     {
05:       Config config = new Config();
06:       config.put("TC_config", args[0]);
07:       TopologyBuilder builder = new TopologyBuilder();
08:       builder.setSpout("TC_gen",
09:                        new TC_Generator());
10:       builder.setBolt("TC_split",
11:                        new TC_Splitter()).
12:       shuffleGrouping("TC_gen"),2);
13:       builder.setBolt("counter",
14:         new TC_Counter()).
15:         shuffleGrouping("TC_split");
16:       Cluster tc_cluster = new Cluster();
17:       tc_cluster.submitTopology("TC_topology",
18:                        config,
19:                        builder.createTopology());
20:     }
21:   }
```
Listing 4: Time-Critical stream application, in charge of allocating the topology and sending it to the cluster.

### B. Map-Reduce Example

Complementing the previous example, this section introduces an example of a time-critical application for Spark (see Figure 6). Our basic example is part of a word-cloud example that reads information from a file to create a word-cloud using Spark map-reduce engine, which is a simple but effective example to illustrate how the modified Spark performs its internal behavior.

From a technical perspective, the chosen analytic loads a file into an RDD (Resilient Distributed Data) and processes it with a simple map function to split input sentences into words, which are latter reduced. The application code is given in Listing 5. To implement the analytic proposed, an RDD is defined to read data (Line 9 of Listing 5). This data is then processed to create a map with all words (Line 11), which is reduced to elements with only one key (Line 13). Lastly, the example stores the resulting data into a file (Line 15).

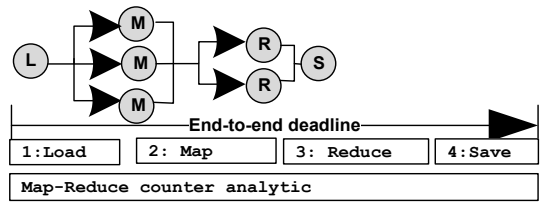

Fig. 6. Structure of the application defined in Listing 5.

```
01: #Simple time-critical work cloud analytic
02: def analytic_map_code (word):
03:        tc_config()
04:        (word, 1)
05: def analytic_reduce_code (a, b):
06:        tc_config()
07:        a+b
08: tc_config()
09: text_file = sc.textFile("file.in")
10: counts = text_file\
11:        .flatMap(lambda line: split(" ")) \
12:        .map(analytic_map_code) \
13:        .reduceByKey(analytic_reduce_code)
14: tc_config()
15: counts.saveAsTextFile("file.out")
```
Listing 5: Time-critical map-reduce for a Spark-Python stack. The example corresponds to a word-cloud that can be configured with the priorities and configuration enforced with `tc_config`.

To ensure time-critical behavior, applications may resort to specific methods that control resource assignment. As in the previous case, there is a `tc_config` specific method, which is charge of enforcing time-critical configuration parameters (defined by the application designer) at the different stages of an application. Depending on the nature of the application, it may delay the application until the next activation, change priorities, or perform any other type of specific optimization.

In Spark, the number of parallel execution units is calculated using the amount of data processed and the number of execution units available. This type of policy is also compatible with the `tc_config` function, because the strategy offers support to all different stages of the model.

## VI. EVALUATION

The basic goals of the empirical section are the following:
- Establish the number of nodes required to satisfy certain required demands, empirically. That it is to provide performance patterns.
- Compare the technique with others, in the cases where this is feasible.
- Establish empirical evidence on the differences offered by a time-critical infrastructure and a general purpose one.

Our evaluation contains a time-critical stack which offers support for a time-critical version of Apache Spark (tc-1.6) map-reduce and Apache Storm (tc-1.6) (see Figure 7 and Table I). Each of these technologies runs on specific time-critical clusters. Lastly, there is a time-critical operating system hosting this entire stack, which runs on a cluster with 60(x4 cores) machines. Those machines share two different storage units: a NFS Linux filesystem used to store data, and a large HDFS engine. The specific versions of Storm modified were rt-0.98 and rt-1.6, which have been extended with time-critical stacks. Our experimental results run until obtain an 1% confidence interval, with a 10-E-10 probability error distribution.

On top of the stack, we run two type of analytics:
- Micro-blogging analytics that count trends. For testing purposes, the evaluation deals with a larger number of tweets processed using offline and online engines. For those applications, the deadline has been taken into account. In this case, the basic formulism used is Eq. 11, which assumes that (B+T=D). It has been iteratively used to calculate number of machines necessary to support a certain time-critical performance.
- Text mining (offline and online). For testing purposes, the evaluation deals with books taken from Project Guttenberg [43] online library. In all cases, the studied analytic counts words to build a histogram.

For text mining and microblogging, the application defines application deadlines for each analytic. Typically, online analytics run with deadlines under one second (<1s) and are supported with the Time-Critical Storm; the offline application has a hard deadline of minutes or hours and are targeted to the Time-Critical Spark (see Table I).

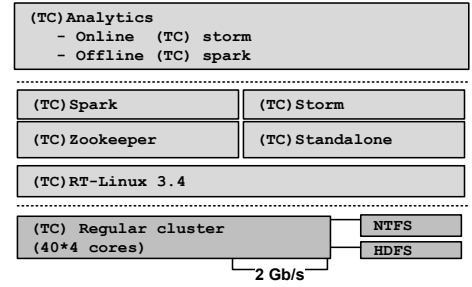

Fig. 7. Type of evaluated map-reduce analytics

Table I: Main parameters of the experiment

|  | |
|---|---|
| Analytics | TC_A1) Micro blogging trending topics (online) <br> TC_A2) Micro blogging word cloud (offline) <br> TC_A3) Word counter (online) <br> TC_A4) Word histogram (offline) |
| Analytic Deadlines | D_TC_A1) Processing an event (< 1 second) <br> D_TC_A2) Processing all data (< 1 hour) <br> D_TC_A3) Processing an event (<1 second) <br> D_TC_A4) Processing all data (<10 min) |
| Confidence interval | +-1% with less than 10E-10 error |
| (TC)Storm version | tc-0.98 |
| (TC)Spark | tc-1.6 |
| Data-set | 1 Terabyte |
| RT-OS | RT-Linux |
| Machines | 240 cores= 60 machines x4 cores per machine |
| Optical Network speed | 1 Gb/s |

### A. Micro-blogging Experiment

This first experiment runs on TC-Storm and it is based on a typical micro-blogging application that counts the number of messages (e.g. tweets) that arrived. In the first case, the application is similar to the one described in Section V for stream processing and map-reduce fluxes.

#### 1) Micro-blogging topics (online)

For micro-blogging, the distributed stream processor selects the most-popular topics which are aggregated to produce an output flux (see Fig. 8). Each different type of stage derives a worst-case computation time (see Table II). For increasing input data frequencies from 1 Hz to 24 kHz, the evaluation analyses the number of resources required to implement the system for the worst-case in which all data require maximum computational times (basically using utilization equations, described from eq. 10 to eq. 11). Also, it was evaluated how the response time is affected by decimation in the aggregation. There is a deadline of 1 second for the end-to-end response time of the stream.

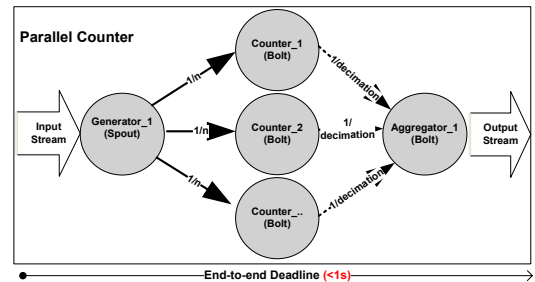

Fig. 8. Type of evaluated multi-input and output stream processor. End to-end deadline comprises from the input to the output

For the given frequencies, the system requires a utilization which may be for the best case satisfied with a single core (e.g. 1 message per second) to a maximum of 83 computational units (see Figure 9, and Figure 10). Figure 9 shows the evolution of this parameter as the number of messages increases. It also shows the number of cores required for an analytic. Using the formalism included in the analytical model, one may establish the end-to-end response time. In all cases, the worst-case observed response time is bounded by 3.4 milliseconds, assuming that there is no decimation.

Previous approaches in the state of the art for Apache Storm only deal with quality of service issues [36]. One major difference among the strategy they proposed in [36] and ours is the type of modeling used. Although the techniques are different (their scheduling techniques) we compared their techniques against ours to determine the number of cores required for their technique. The results showed that their performance takes more from 22 % to 33% (depending on the scenario) because their equations do not take into account blockings factors (B) and Eq. 8 and Eq. 11 are over-pessimistic (mainly because in modeling they do not model blockings).

Table II: Main parameters of the experiment. Partial costs of the three stages and number of cores required to implement the "time critical" system.

| Parameter | Value |
|---|---|
| Costs ($C_{gen}$, $C_{counter}$, $C_{aggreg}$) | 127 µs, 507 µs, 511µs |
| Data In freq | 1 Hz -4 kHz |
| Cores Available | 1-28 |
| End-to-end Deadline | <1 second |
| TC stack required resources to meet the deadline | From for 1(0,03) core for 1Hz – 80 cores for 4 kHz |
| Closer approaches [36] | From for 1(0,04) core for 1Hz – 104 cores for 4 kHz |

From the point of view of performance, the sampling factor is relevant, because it may reduce the number of cores required to implement the system. In our particular case, a proper decimation reduces the amount of cores assigned to the aggregator phase, increasing the response time of the analytic too (because of the delay introduced in the application end-to-end deadline). In the experiment, the end-to-end response time increases from few milliseconds to a second (see Figure 11) because of the decimation factor. For the same type of applications, the proposed architecture reduces the aggregator costs by delaying the transmission of data, which results in reductions on the number of cores required to submit data among nodes (Figure 12). Figure 12 refers to the savings associated with the use of a decimation technique that reduces the number of data sent from the client to the server. Figure 12 refers to the number of resources one saves as one does not have to send data. For a decimation factor of 2000, the end-to-end response is below one millisecond, saving for the 1 kHz a maximum of 2 extra resources (i.e. cores).

Figure 9 shows the utilization required to implement each section of the system. This factor depends on two input parameters (the stream input frequency: from 0 to 20 MHz and also the type of segment: which may be the generator, the counter and the aggregator). The term refers to the partial contribution of each segment (C/T) of the stream, included in Eq. 11. Taking the input information of Figure 9, Figure 10 shows to the minimum number of resources (i.e.) decomposed by partial contributions (of the generator, counter, and aggregator) for different frequencies. The difference among both figures is that first refers to utilization and the second to an integer number of cores.

The common goal of Figures 11 and 12 is to illustrate the benefits of decimation among phases to save resources, which is also an important parameter to take into account. In Figure 11 to introduce the impact of the decimation the previous analysis has to be changed introducing a delay or blocking in the end-to-end deadlines. As a result of this change, a number of resources are released that depend on the decimation factor used (from 1 to 1024) and the input frequency (from 100 Hz to 16 KHz). This saves a number of cores (shown in Figure 12) that may be relevant.

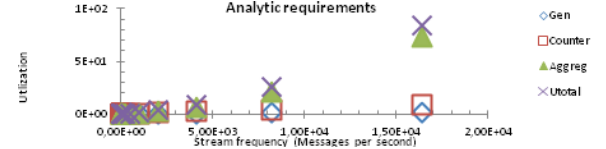
Fig. 9. Analytic demanded resources (Utilization)

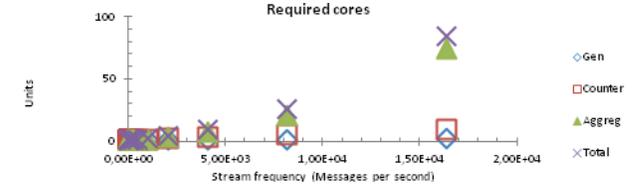
Fig. 10. Required number of cores to support the scenario

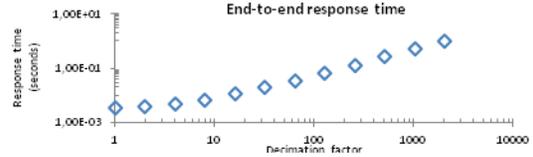
Fig. 11. End-to-end response time as a function of a decimation (i.e. sampling in the output) factor

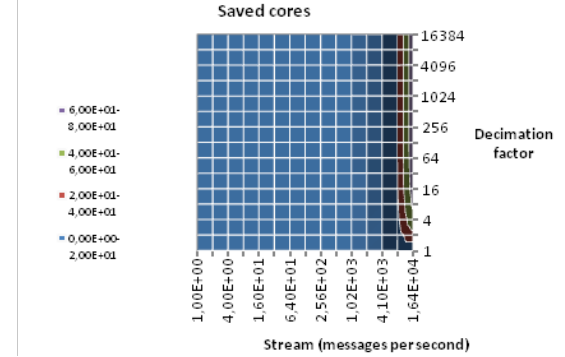
Fig. 12. Saved cores as the output of the counters are delayed to meet a 1 second deadline. x axis refers to the speed of the microblogging flux.

*2) Time-critical Offline performance*

Our second type of analytic on micro-blogging applications is the processing of splitting and processing a set of tweets (see Figure 13). Our application processes a large amount of data stored in a HDFS filesystem to obtain a word-cloud. The data consists of four main stages, one that downloads from the HDFS, another which performs a map to tokenize data, a reduce phase which groups by similar words, and a final stage which sorts them (see characterization is in Table III and Figure 13). The application dataset processes data from 160 millions of words to 1300 millions of words that may be processed with 128, 64, 32, 16, 8, 4, 2, and 1 cores. There is also a deadline for the analytic of 2 hours, which requires a minimum of 20 cores to be satisfied all scenarios. This formulism has been derived from Eq. 11 if we assume (T+B=D).

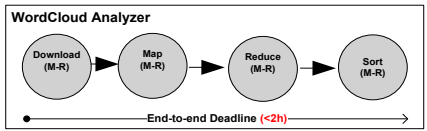

Fig. 13. Wordcloud analysis over a micro-blogging application. Analytic that consists of 4 stages with an end-to-end deadline of 2 hours.

As in the previous experiment, we established empirical evidence to compare our scheduling model against results from a map-reduce framework [11]. Previous work does not consider the use of blockings in the model leaving to very pessimistic performance that may be improved adding new rules to [11] (see Table III). This is mainly due to the blockings caused to access data from the Hadoop distributed file-system. The technique described in [11] requires from 25% to 33% of extra resources for the analyzed scenarios. Table III shows the parameters of the analytic, and the amount of resources required in each case.

Table III: Main parameters of the experiment. Main results

| Parameter | Value |
|---|---|
| Analytic stages | download() -> map () -> reduce() -> sort() |
| Data | HDFS 1, 2, 4, 8 GB of data (tweets) |
| Cores | 1-128 |
| Memory per core | 1 GB |
| Partitions of Spark | 300 |
| Analytic Deadline | < 2 hours |
| Our time-critical infrastructure | It requires 1 (0,5) core for 1 GB of data and 20 cores for 8 GBs to meet the deadline |
| Closest related work [11] | Requires (0,76) cores for 1 GB and 28 for 8 GB of data |

Figure 14 describes all the experiments carried out. It includes the number of cores used in the experiment (diamonds) and also the data (squares). Each point in the x-axis represents one experiment (which consists of a number of cores and data that has to be processed labeled as UNITs). Figure 15 extends Figure 14 with the time taken to carry out the whole experiment. Each scenario (cores and data) produces an output (triangle). It also includes the description for the speed of the scenario (speed means the amount of time delivered by each core) and efficiency.

The following performance patterns have been observed:
- *Time:* The total time required to run each experiment has a relationship with the amount of processed data (more data means more time). Also, it also decreases as the number of cores increases (see Figure 15). To meet the deadline, the system requires 20 cores: with 16 cores, it takes 2.2 hours to compute the largest file; and with 32 cores, it takes 1.6 hours.
- *Speed*: The speed, measured as the number of tweets per second, decreases as the number of cores does. It also has dependency on the data transferred but the main dependency is with the number of cores available to process data (Figure 16).
- *Efficiency*: The efficiency measured as the speed divided by the number of cores required to implement a system increases as the number of cores decreases (Figure 17).

In general, an increase in the number of nodes does not mean more speed or efficiency. This is because of the overhead introduced by connections. In the proposed analytic, the bottleneck is the network. This is reason why adding cores does not linearly increase the speed of the application.

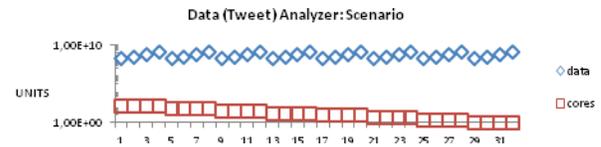

Fig. 14. Offline micro-blogging scenario

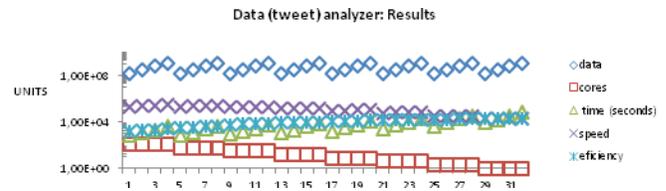

Fig. 15. Off-line micro-blogging scenario: results

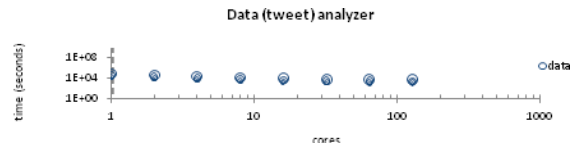

Fig. 16. Total time results

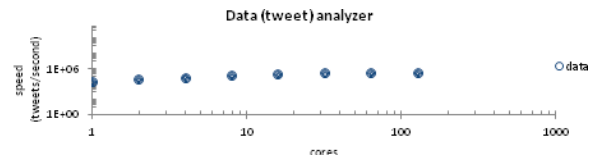

Fig. 17. Speed results

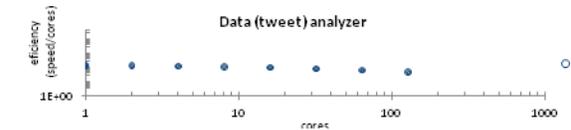

Fig. 18. Efficiency results

*B. Book Manager*

The second evaluation analytic is the book manager, which operates on the Gutenberg project [44]. As in the previous case, it consists of two subsystems: one in charge of processing sentences of a book online, and another which processes a set of books offline which are accessed from HDFS. The goal is the same as in the previous section: to establish empirical evidence on the performance one may expect from this type of infrastructures. As in the previous case, our interval of confidence is 1% with a failure probability of 1e-10.

*1) Book word histogram (online)*

The structure of the analytic is shared with the micro-blogging analytic (shown in Fig 13). Also the end-to-end analytic is one second. However, the worst-case computational models for the different stages are higher than in the previous case (see Table IV). In the previous case, all stages were under the millisecond response time, now some of them are close to 5 ms. As there is an increase in the computation time in all stages, the number of nodes required to implement the system is also higher. Figure 19 and Figure 20 introduce the costs in utilization demanded by the application and number of cores required to meet the deadline. They are always greater than in the micro-blogging application. Likewise, it is expected that decimation increases the end-to-end response times of the analytic but it also reduces the amount of resources required to be implemented in the cluster.

Table IV: Main parameters of the experiment and outcomes

| Parameter | Value |
|---|---|
| Costs: $C_{gen}$, $C_{counter}$, $C_{aggreg.}$ | 1,1 ms, 5 ms, 0,8 ms |
| Data Input freq | 1 Hz -40 kHz |
| Cores available | 1-128 |
| Analytic Deadline | <1 second |
| Our time-critical infrastructure | Requires 1 (0,08) core for 1 Hz and 100 cores for 40 kHz |
| Closest related work [32] performance | Requires 1(0,096) core for 1 Hz and 119 cores for 40 kHz |

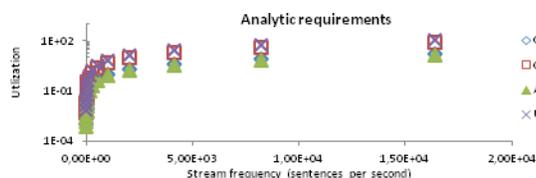
Fig. 19. Analytic Requirements (Utilization)

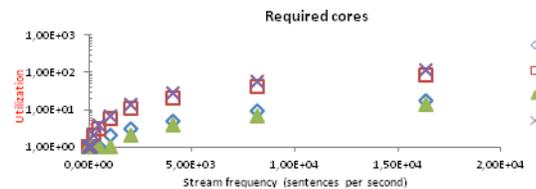
Fig. 20. Number of cores required for the implementation

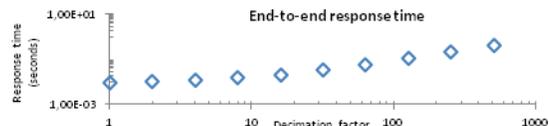
Fig. 21. Response time with decimation

*2) Book word histogram (offline)*

As in the previous case, the offline library analyzer splits the book into sentences and words. The used test-bed (see Table V) ranges from 1 book (6.9 MB) to 512 books (3.1 GB). The number of cores required to support the system ranges from 1 to 50. Internally, Spark creates 200 partitions. The number of items to be processed ranges from 1 million elements to 1000 millions. The number of cores also ranges from 1 to 50 core(s) (see Fig. 21). Results of the experiment are shown in Fig. 22. They have been scheduled using the model proposed in Eq. 11 (T+B=D).

The analysis of the results shows that the processing time has a strong relationship with the amount of processed data (Fig. 23). The speed keeps more or less stable in all experiments (Fig. 24). Likewise, the efficiency tends to be higher with lower number of machines and decreases as the number of machines increases (Fig. 25).

To meet the deadline for the whole analytic (T+B=D), the system requires at least two cores (which process the system in 9.79 minutes). With one core, the response-time is 13 minutes and with 2 cores is 9.70 minutes. Comparison with similar techniques (i.e. [11]) showed that it may require a 40% of extra resources to meet deadlines (see Table V).

Table V: Main parameters of the experiment. Results

| Parameter | Value |
|---|---|
| Analytic stages | `download()-> map ()->reduce()->sort()` |
| HDFS Data | From: 1 book (6,9 MB) to: 512 books (3.1 GB) |
| Cores | 1-50 |
| Partitions of Spark | 200 |
| Analytic Deadline | <10 min |
| Our time-critical infrastructure | From 1 (0,013) core for 1 book to 2 cores for 512 books |
| Closest related work [11] performance | From 1 (0,027) core for 1 book to 4 cores for 512 books |

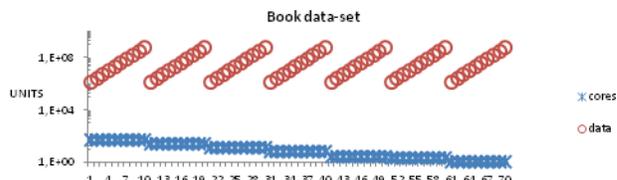
Fig. 22. Book word histogram generator.

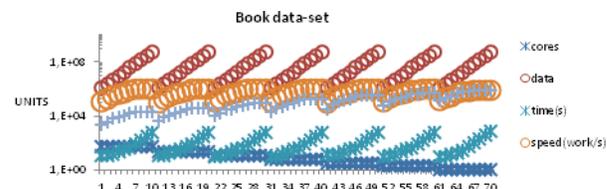
Fig. 23. Book word histogram

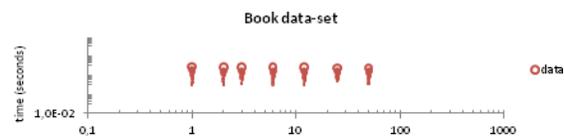
Fig. 24. Detailed results on total time

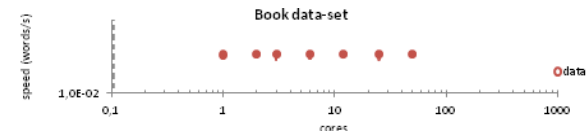
Fig. 25. Detailed results on speed

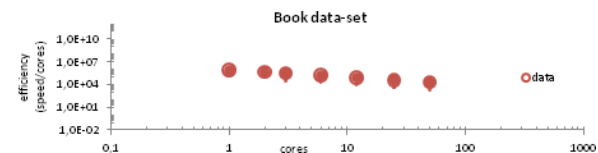
Fig. 26. Detailed results on efficiency

*C. General Purpose vs. Time Critical Performance*

To illustrate a case where the time critical information is used, a simple example is presented. It is based in the idea of prioritization offered by the infrastructure. Let's assume that we have two time-critical analytics ($TC_1$ and $TC_2$). Each one of them takes computationally 1 hour to run ($C_{TC1}=C_{TC2}=1$ hour). But these time critical analytics have different deadlines ($D_{TC1}=1$ hour and $D_{TC2}=2$ hours). If we do not assume any time-critical scheduling model (like the one shown in Section V), then the system is not feasible in a single node; it requires two, because both tasks have the same priority ($P_{TC1}=P_{TC2}=prio$), which it is the default configuration in Spark and Storm. With this setup, the worst case response time of both map-reduce tasks is 2 hours ($RT_{TC1}=RT_{TC20}=2$ hours) requiring an extra machine to isolate applications with different deadlines. Here the formulism used to calculate worst case computation times is derived from Eq. 9. However, our scheduling framework may assign priority proportionally to deadlines ($P_{TC1}$=prio and $P_{TC2}$=prio+1), requiring only one resource to meet deadlines (see Table VI). In this case, the worst-case for the shortest deadline reduces to 1 hour, and one single machine, the worst case response time for the highest priority task does not suffer interference from the lowest one.

Table VI: Cluster with one machine allocating time critical (TC) and general purpose (GP) analytics. TC cluster is much more efficient deadline with deadlines.

| | | Cost | Deadline | Priority | WCRT | Feasible |
|---|---|---|---|---|---|---|
| GP | TC_1 | 1 h | 2h | Default | 2h | Y |
| | TC_2 | 1 h | 1h | Default | 2h | **N** |
| TC | TC_1 | 1 h | 2h | Low | 2h | Y |
| | TC_2 | 1 h | 1h | High | 1h | Y |

## VII. RELATED WORK

In the state-of-the-art of time-critical big-data systems, different approaches have been identified as pioneering efforts that contribute to sculpt the time-critical big-data infrastructures. Each one contributes from a different perspective to different aspects of next-generation architectures for big-data. Chronologically, the first is an attempt to model real-time map-reduce interactions as schedulable entities [11]. In [11] the authors used the popular Hadoop map-reduce model which has been evaluated on an experimental Amazon EC2 cloud, establishing interesting tradeoffs between throughput and predictability. The described model may be improved with the blocking formulism described in Section IV, as our empirical results suggested.

From the point of view of time-critical systems, this is one of the first approaches to deal with the time-critical performance of map-reduce applications. Later, the Juniper project [10] has dealt with a number of issues related to the performance of big-data systems.

More recently, some researchers [9] have addressed the predictability of Apache Storm, one of the main online infrastructures available for stream processing, as part of an all-in-Java infrastructure for real-time big-data. To this end, they introduced programming abstractions typically used in time-critical systems into the Apache Storm architecture [9]. Their major contribution to the state-of-the-art has been the integration of a computation model based on stream processing with the scheduling policies available for distributed and parallel computing.

A recent approach to on-line processing is the lambda architecture [1], which is based on a batch technology and online technology that provides a dual branch computing model. The architecture proposed in this section enables the possibility of defining requirements (typically a deadline) for the analytics, which are efficiently enforced latter by the infrastructure. This is also more efficient than the performance given by current infrastructure, which targets to high performance computing. Our particular time-critical lambda architecture will consist of a TC Storm for stream processing and a TC-Spark for batch processing. The contribution of the article to the lambda architecture is to be able to use the real-time scheduling theory to derive efficient end-to-end scheduling models from deadlines.

### A. Distributed Stream Processing

In the area of distributed stream processing, there are a number of initiatives dealing with distributed stream processing [35-39]. The work in [35] introduces QoS scheduling mechanisms for Apache Storm. In [36] the authors added adaptive scheduling techniques to Storm. Our approach uses the techniques described in [35-36] to offer a deadline-based approach, which is a domain not addressed by previous researchers. Our scheduling model shares commonalities [36] that may benefit from blockings factor introduced to improve significantly the schedulability of the system.

Some other approaches [37-38] deal with scheduling models for clusters and the cloud. The main difference among these two techniques and the proposed technique is the domain. While those techniques address general purpose scheduling stochastic models, our approach deals with worst-case analytics which offer a simpler formalism.

### B. Map-Reduce Processing

There is a corpus of works [39-42][48-49] dealing with different aspects involved in map-reduce. In [40], the authors describe different quality-of-service features related to a map-reduce engine. Our algorithms belong to the response time category of the quality of service. In [39], the authors formalize aspects in map-reduce scheduling to perform online and offline scheduling. Our model is much simpler than [39]. In [41] the authors proposed a packing server for map-reduce workflows. We share with this work a packing strategy; however, we explicitly split flushes into different units. Lastly, [42] describes a cost-effective scheduling framework for map-reduce. The framework takes into account monetary issues. Our main difference [41-42] is that our approach is more empirical, targeting specific end-systems.

In [48], an architecture is proposed for high-speed performance based on RabbitMQ. The system described in [48] implements a scheduler able to run analytics. The main difference among both approaches is the domain, which in our case it is more focused on the use of time-critical scheduler and applications. A similar assessment may be applied to [49], which does not define polices for deadline processing. A last piece of work in the map-reduce world is described in [51] that uses scheduling servers with map-reduce tasks. Since both use different scheduling models, the techniques cannot be easily compared.

## VIII. CONCLUSIONS AND FUTURE WORK

Many challenges are ahead of us in the time-critical big-data horizon towards producing a generic infrastructure able to meet the deadlines of different analytics in a predictable way. This article has reviewed some (small) building blocks that have to be considered to accomplish this goal; it has also identified requirements of different domains that have to be properly supported by big-data infrastructures and must be readapted to cope with time-critical issues.

Among all of these building blocks, one of increasing importance is big-data analytics and their particular characteristics, which may determine the type of required infrastructure. Only with careful consideration of the characteristics of the different types of analytics, will it be possible to unravel the requirements of next-generation time-critical big-data platforms.

Currently, the authors are considering the integration of privacy and security; where they are focusing their efforts in studying the overhead introduced by different authentication policies, partially described in [19], as a part of a time-critical big-data system.

ACKNOWLEDGEMENTS

This work been partially supported by the Spanish National Education Ministry, under Jose Castillejo Program: Infrastructure for Real-Time Big-Data (CAS14/00118), and the national program: HERMES-SMARTDRIVER (TIN2013-46801-C4-2-R) and AUDACity (TIN2016-77158-C4-1-R). It has been also partially funded by European Union's 7$^{th}$ Framework Programme under Grant Agreement FP7-IC6-318763 and eMadrid (S2013/ICE-2715). Some experiments presented in article were inspired in the Grid'5000 testbed, supported by a scientific interest group hosted by Inria and including CNRS, RENATER and several Universities as well as other organizations (see https://www.grid5000.fr). We are also in debt with our anonymous reviewers that provided valuable feedback to improve the article.

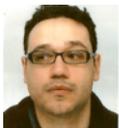 **Pablo Basanta-Val** was born in O Valadouro, Lugo, Spain. He received the Telecommunication Engineering degree from the Universidad de Vigo, Spain, in 2001 Madrid, Spain, in 2007. Currently, he is an Associate Professor at the Telematics Engineering Department, Universidad Carlos III de Madrid and the UC3M-BS Institute of Financial Big Data (IFiBiD). He was member of the Distributed Real-Time Systems Lab from 2001 and since 2014 is enrolled in the Web Semantic Lab. His research interests are in Real-Time Java technology and general-purpose middleware used to support next generation applications. His current interests are on the definition and evaluation of predictable models for big data infrastructures. He has authored/co-authored over 90 papers/reports. He participated in a number of European projects such as ARTIST, ARTIST NoE, ARTIST Design, iLAND and several national projects.

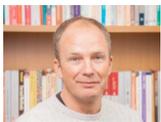 Professor **Neil Audsley** received a BSc (1984) and PhD (1993) from the Department of Computer Science at the University of York, UK. In 2013 he received a Personal Chair from the University of York, where he leads a substantial team researching Real-Time Embedded Systems. Specific areas of research include high performance real-time systems (including aspects of big data); real-time operating systems and their acceleration on FPGAs; real-time architectures, specifically memory hierarchies, Network-on-Chip and heterogeneous systems; scheduling, timing analysis and worst-case execution time; model-driven development. Professor Audsley's research has been funded by a number of national (EPSRC) and european (EU) grants, including TEMPO, eMuCo, ToucHMore, MADES, JEOPARD, JUNIPER, T-CREST and DreamCloud. He has published widely, having upwards of 150 publications in peer reviewed journals, conferences and books.

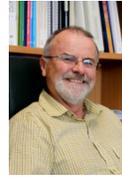 **Andy Wellings** is Professor of Real-Time Systems at the University of York, UK in the Computer Science Department. He is interested in most aspects of the design and implementation of real-time dependable computer systems and, in particular, real-time programming languages and operating systems. He is European Editor-in-Chief for the Computer Science journal "Software-Practice and Experience" and a member of the International Expert Groups currently developing extensions to the Java platform for real-time, safety critical and distributed programming. Professor Wellings has authored/co-authored over 300 papers/reports. He is also the author/co-author of several books including "Concurrent and Real-Time Programming in Ada", "Real-Time Systems and Programming Languages (4th Edition)" and "Concurrent and Real-Time Programming in Java".

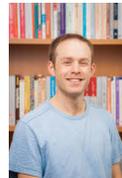 **Ian Gray** is a Research Fellow at the University of York, UK. His Ph.D., which was completed at York, concerned the use of novel virtualisation techniques to overcome the limitations of traditional programming models for the development of complex embedded systems. This work was further developed as part of the EU-funded MADES project. He has been part of three successful FP7 research projects, author of many papers and book chapters, and programme committee member of EuroMPI and MCSoC. Ian's work is centered around real-time, embedded systems, and particularly the use of new programming models and techniques to help the development of systems with complex hardware architectures.

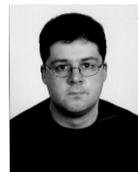 **Norberto Fernández-García** received his Telecommunication engineering degree from Universidad de Vigo (UVigo), Spain, in 2002, and the Ph.D. degree from Universidad Carlos III de Madrid (UC3M), Madrid, Spain, in 2007. Currently he works as a Ph.D. teaching assistant at the Defense University Center located at the Spanish Naval Academy. He has carried out most of his research activity in the area of networked information systems, working on the application of artificial intelligence techniques and tools to information management (Semantic Web / Linked data) and, more recently, in the development of techniques to efficiently process large volumes of data (Big data). He has been involved in 15 research projects at regional, national and international levels. He has also co-authored more than 40 publications, including 20 publications in national and international journals.